\documentclass[a4paper,11pt]{article}
\pdfoutput=1 
\usepackage{amsmath}
\usepackage{graphicx}
\usepackage{float}
\usepackage{mathrsfs,graphicx,rotating,amsmath,amsfonts,mathtools,booktabs,wasysym,caption}
\usepackage{slashed}
\usepackage[table,xcdraw,dvipsnames]{xcolor}
\usepackage{graphicx}
\usepackage{bbold}
\usepackage[utf8x]{inputenc}
\usepackage[english]{babel}
\usepackage{multirow,multicol}
\usepackage{epstopdf}
\usepackage{bbm}
\usepackage{changepage}
\usepackage{appendix}
\usepackage{libertine}
\usepackage{braket}
\usepackage{wasysym}
\usepackage{empheq}
\usepackage{cancel}
\usepackage{enumitem}
\usepackage{mathrsfs}
\usepackage{lmodern}
\usepackage{tabularx}
\usepackage{multicol}
\usepackage{color}
\usepackage{mathtools}
\usepackage{verbatim}
\usepackage{amssymb}
\usepackage{amsfonts}
\usepackage{jheppub} 

\usepackage[T1]{fontenc} 

\title{Holographic Perspectives On Models Of Moduli Stabilization In M-Theory}

\author[a]{Sirui Ning}

\affiliation[a]{Rudolf Peierls Centre for Theoretical Physics
Beecroft Building, Clarendon Laboratory, Parks Road, University of Oxford, OX1 3PU, UK}

\emailAdd{sirui.ning@physics.ox.ac.uk}

\abstract{Recent holographic analyses on IIA and IIB models of moduli stabilization have led to many interesting results. Here we extend this approach to M-Theory. We consider both flux-stabilized models and non-perturbative stabilization methods. We perform a holographic analysis to determine the spectrum of the assumed dual $CFT_{3}$ to see its AdS/CFT implication. For the flux stabilization,  which relies on a large complex Chern-Simons invariant, moduli have integer dimensions similar to the DGKT flux-stabilized model in type IIA,. For the non-perturbative stabilization, the results are similar to racetrack models in type IIB.}

\begin{document} 
\maketitle
\flushbottom
\section{Introduction}
The properties of the vacuum in string theory are important because if string theory is true, then the vacuum will be able to describe our universe. The approach to connect the original formulation of string theory and the real world is string compactification (for reviews see \cite{Conlon:2006gv, Douglas:2006es,Denef:2008wq,Denef:2007pq}), which reduces the full 10d string vacuum to a (3+1)-d spacetime, while other dimensions are compactified. However, in the process of string compactification, new massless scalar fields will emerge. The vacuum expectation values of them, called moduli, do not at first appear in the 4d effective potential, so they are not constrained. 

If moduli can take arbitrary values (for example, time dependent), then this will conflict with observations. Therefore, it is essential to stabilize the moduli. The ingredients of moduli stabilization are fluxes and non-perturbative effects (for example, world sheet instantons \cite{Dine:1986zy} and gaugino condensation) to make new terms appear in the 4d effective potential to minimize the moduli. Fluxes can be seen as a higher dimensional generalization of Dirac quantization, so they take integer values. Therefore, the corresponding vacua are also labeled by discrete integer fluxes. The set of all vacua generated by these fluxes form the so-called "landscape".

Examples of scenarios of moduli stabilization include the type IIA string compactifications like DGKT \cite{DeWolfe:2005uu} and type IIB string compactifications like the Large Volume Scenario(LVS)\cite{Balasubramanian:2005zx,Conlon:2005ki,Conlon:2006wz,Cicoli:2007xp}, KKLT \cite{Kachru:2003aw} and Racetrack \cite{Escoda:2003fa}. In DGKT all geometric moduli are stabilized at SUSY AdS vacua in the large flux limit. In the Large Volume Scenario the moduli are stabilized at non-SUSY AdS vacua with an exponentially large volume. KKLT stabilizes the moduli with both fluxes and non-perturbative effects at SUSY vacua in the limit of small $W_{0}$. The Racetrack model uses two different non-perturbative effects to stabilize the moduli at a SUSY vacuum without fluxes.

M-Theory is another limit of string theory. Moduli stabilization is also interesting for M-Theory compactification. This has been studied in \cite{PhysRevLett.97.191601,Acharya:2002kv,Acharya:2007rc} by both flux and non-perturbative stabilization. In the flux-stablized vacua, all the saxions and a linear combination of axions are stabilized at SUSY AdS vacua under certain topological conditions. The non-perturbative stabilized vacuum is the multiple moduli version of the racetrack model.

The above method goes from top to down. However, in recent years an alternative approach has risen: the swampland program \cite{Palti:2019pca,Ooguri:2006in,Vafa:2005ui}. This aims at using basic principles in quantum gravity to exclude many effective field theories and work out properties that low energy effective field theory must satisfy. It is beneficial to understand the swampland program from holographic perspectives: for references see \cite{Benjamin:2016fhe,Nakayama:2015hga,Montero:2016tif,Giombi:2017mxl,Urbano:2018kax,Harlow:2018tng,Harlow:2018jwu,Montero:2018fns,Baume:2020dqd,Perlmutter:2020buo}.

A novel approach from holography was proposed \cite{Conlon:2018vov,Apers:2022zjx,Conlon:2020wmc,Conlon:2021cjk,Apers:2022tfm,Quirant:2022fpn}, which provides a new perspective to the problem. The motivation is to determine the consistency of these 4d vacua from the CFT side. So far, what has been studied are the LVS, the fibred LVS and DGKT. In particular DGKT gives an interesting spectrum leading to the integer conformal dimensions.
  
In this paper we extend the holographic swampland story to M-Theory, both models with fluxes and without fluxes. The paper is organised as follows. In Section 2 we describe general aspects of M-Theory moduli stabilization on a $G_{2}$-holomony manifold and the mass matrix elements. In Section 3 we study the holographic dual of M-Theory vacuum with flux and compare it with the previous results of DGKT \cite{Conlon:2021cjk,Apers:2022tfm}.  In Section 4 we study the holographic dual of M-Theory vacuum with zero flux background and compare it with the previous results of KKLT and racetrack \cite{Conlon:2020wmc}. In Section 5 we give our conclusions. In the Appendix the detailed calculations are presented.
\section{General Aspects Of M-Theory Moduli Stabilization}
In this section we give a brief description of M-Theory moduli stabilization on a $G_{2}$ holonomy manifold \cite{Acharya:2002kv}. The 11d low energy supergravity description of M-theory is given by the following action \cite{Cremmer:1978km}:
\begin{equation}
S=\frac{1}{2 \textit{k}_{11}^{2}}\left[\int d^{11}x\sqrt{-g}R-\int \left(\frac{1}{2}G_{4}\wedge \ast G_{4}-\frac{1}{6} C_{3}\wedge G_{4}\wedge G_{4}\right) \right],
\end{equation}
where the 11d metric $\textit{g}$ and a 3-form $C_{3}$ consists of the bosonic components. $G_{4}=dC_{3}$ is the field strength.
By compactifying this theory on a 7d compact manifold X with $G_{2}$ holonomy group, one obtains the 4d $\mathcal{N}=1$ supergravity theory. A covariantly constant 3-form $\phi$ always accompanies a manifold X with $G_{2}$ holonomy. Following the argument in \cite{Acharya:2002kv}, the moduli space of X has the same dimension as $H^{3}(X,R)$, therefore the 4d $\mathcal{N}=1$ supergravity theory has $b_{3}(X)$ moduli.
The corresponding chiral fields read: 
\begin{equation}
z_{i}=t_{i}+i s_{i},\quad i=1,...N,
\end{equation}
$s_{i}$ are volume of 3-cycles and $t_{i}$ are the corresponding axions. 
In the case that is consistent with $G_{2}$ holonomy \cite{Acharya:2007rc}, the K\" ahler potential is expressed by:
\begin{equation}
K=-3\textrm{Log}(\mathcal{V}).
\end{equation}
We assume, following \cite{Acharya:2007rc},  the volume of the 7d-manifold can be written as $\mathcal{V}=\prod_{i=1}^{N}s_{i}^{a_{i}}.$
The parameters have the following constraints:
\begin{equation}
\Sigma_{i=1}^{N}a_{i}=\frac{7}{3}.
\end{equation}
The K\" ahler potential satisfies the following no scale relationships:
\begin{equation}
\label{noscale}
K^{ij}K_{j}=-s^{i}, \quad K^{ij}K_{i}K_{j}=7.
\end{equation}
We now derive general expressions for the Hessians of both volume moduli and axions, assuming that we have a SUSY vacuum. Details are presented in  Appendix \ref{a1}. We start from the standard $\mathcal{N}=1$ supergravity potential form (taking $M_{P}=1$):
\begin{equation}
\textrm{V}=e^{K}(K^{i\overline{j}}D_{i}WD_{\overline{j}}\overline{W}-3|W|^{2}).
\end{equation}
The SUSY condition is:
\begin{equation}
\label{eq:susy}
D_{i}W=\partial_{i}W+K_{i}W=0.
\end{equation}
For the volume moduli, the corresponding Hessian is expressed:
\begin{equation}
\begin{aligned}
\label{eq:mass matrix}
\partial_{b}\partial_{a}\textrm{V}&=K_{ab}\textrm{V}-3e^{K}K_{b}\partial_{a}|W|^{2}-3e^{K}\partial_{b}\partial_{a}|W|^{2}+e^{K}K^{i\overline{j}}(\partial_{a}D_{i}W)(\partial_{b}D_{\overline{j}}\overline{W})\\
&+e^{K}K^{i\overline{j}}(\partial_{b}D_{i}W)(\partial_{a}D_{\overline{j}}\overline{W}).\\
\end{aligned}
\end{equation}

Following Appendix \ref{a1} (also see \cite{Apers:2022tfm}),  putting everything together:

\begin{equation}
\label{eq:3}
\begin{aligned}
H^{V}_{ab}=\frac{\textrm{V}_{ab}}{e^{K}|W|^{2}}&=-K_{ab}+3K_{a}K_{b}+2\frac{W_{ab}}{W}+8K^{ij}\frac{W_{ia}W_{jb}}{|W|^{2}}\\
&+2s_{i}K_{b}\frac{W_{ia}}{W}+2s_{j}K_{a}\frac{W_{jb}}{W},\\
\end{aligned}
\end{equation}
where $K$, $W$ are the K\"ahler potential and superpotential evaluated at the SUSY vacuum. 

The Hessian for the axions can be obtained similarly from Eq (\ref{eq:mass matrix}), with all derivatives of the K\"ahler potential vanishing:
\begin{equation}
\begin{aligned}
\partial_{b}\partial_{a}V&=-3e^{K}\partial_{b}\partial_{a}|W|^{2}+e^{K}K^{i\overline{j}}(\partial_{a}D_{i}W)(\partial_{b}D_{\overline{j}}\overline{W})\\
&+e^{K}K^{i\overline{j}}(\partial_{b}D_{i}W)(\partial_{a}D_{\overline{j}}\overline{W}),\\
\end{aligned}
\end{equation}
where $\partial_{a}$,$\partial_{b}$ means $\partial_{t_{a}}$, $\partial_{t_{b}}$.
Again, following Appendix \ref{a1} (also see \cite{Apers:2022tfm}), we can get the axion Hessians similarly:
\begin{equation}
\begin{aligned}
\label{eq:4}
H^{A}_{ab}&=\frac{\partial_{t_{b}}\partial_{t_{a}}V}{e^{K}|W|^{2}}=2K_{a}K_{b}+6\frac{W_{ab}}{W}+8K^{ij}\frac{W_{ia}W_{jb}}{|W|^{2}}\\
&+2s_{i}K_{b}\frac{W_{ia}}{W}+2s_{j}K_{a}\frac{W_{jb}}{W},\\
\end{aligned}
\end{equation}
where  $\partial_{a}$,$\partial_{b}$ means derivatives to  $\partial_{s_{a}}$, $\partial_{s_{b}}$.\\

We now apply this formalism to two models of moduli stabilization. The first is flux-stabilization and the second is non-perturbative stabilization. In practice the superpotential will be either Eq (\ref{super1}) or Eq (\ref{super2}). 
\section{Flux-Stabilized M-Theory Vacuum}
This section investigates the moduli stabilization method of \cite{Acharya:2002kv,Acharya:2005ez}, which aims at stabilizing the moduli with fluxes. This is analogous to the DGKT model in type IIA, which also uses fluxes. We give an introduction to M-Theory compactifications on $G_{2}$ manifolds with fluxes turned on.

In order to stabilize the moduli, certain topological conditions have to be imposed: The $G_{2}$ manifold X needs to have an ADE singularity along a 3-manifold Q \cite{Acharya:2002kv}. Following \cite{Beasley:2002db,Acharya:2002kv}, the superpotential induced by turning on a background flux G and background fields at the singularities is:
\begin{equation}
W=\frac{1}{8\pi^{2}}\int \left(\frac{C_{3}}{2}+i\phi \right)\wedge G+c_{1}+ic_{2}.
\end{equation}
Expanding the 4-form flux G into a harmonic basis: 
\begin{equation}
G=N^{i}\rho_{i},
\end{equation}
where $N_{i}$ are fluxes and $\rho_{i}\in H^{4}(X,\mathbb{Z})$, results in the following superpotential:
\begin{equation}
\label{super1}
W(z)=N^{i}z_{i}+c_{1}+ic_{2}.
\end{equation}
Following \cite{Acharya:2002kv},  $c_{1}+i c_{2}$ is a complex Chern-Simons invariant. This term is essential for this mechanism because without this term the moduli could not be stabilized and the cosmological constant would be zero. Without loss of generality we can take $c_{2}$ to be positive. 

The SUSY condition implies:
\begin{equation}
\frac{-3a_{i}}{2i s_{i}}(N^{j}z_{j}+c_{1}+ic_{2})+N_{i}=0.
\end{equation}
Separating the real and imaginary part of this equation, we have:
\begin{equation}
\begin{aligned}
&N^{i}t_{i}+c_{1}=0,\\
&\frac{3a_{i}}{2s_{i}}(N^{j}s_{j}+c_{2})=N^{i},\\
\end{aligned}
\end{equation}
which results in:
\begin{equation}
s_{i}\frac{3a_{i}}{2s_{i}}(N^{j}s_{j}+c_{2})=\frac{7}{2}(N^{j}s_{j}+c_{2})=N^{i}s_{i}.
\end{equation}
This leads to the following solution:
\begin{equation}
\begin{aligned}
&c_{2}=-\frac{5}{7}N^{i}s_{i},
&s_{i}=-\frac{3a_{i}}{5N_{i}}c_{2},
\end{aligned}
\end{equation}
which implies a vacuum expectation value $W=-i \frac{2c_{2}}{5}$.\\

Before we proceed, we need to check the validity of the effective field theory description. In order to realize scale separation, we need the Kaluza-Klein radius to be much smaller than the AdS radius:
\begin{equation}
\begin{aligned}
&R^{2}_{AdS}=\frac{1}{e^{K}|W|^{2}}=\frac{\mathcal{V}^{3}}{|W|^{2}}\sim \frac{c_{2}^{7}}{c_{2}^{2}}\sim c_{2}^{5},\\
&R^{2}_{KK}\sim \mathcal{V}^{\frac{2}{7}}\sim c_{2}^{\frac{2}{3}}.
\end{aligned}
\end{equation}
so $R_{KK}\ll R_{AdS}$ if $c_{2}\gg 1$. Following \cite{Acharya:2002kv}, we assume that large values of $c_{2}$ can be obtained for some special $G_{2}-$manifolds. 

This superpotential implies that the second order derivatives of the superpotential all vanish. Following Eq (\ref{eq:3}), the Hessian corresponding to the volume moduli is:
\begin{equation}
M_{ij}=-K_{ij}+3K_{i}K_{j}.
\end{equation}

The physical masses can be extracted as the eigenvalues of $m=2K^{-1}M$, where the matrix elements read:
\begin{equation}
\begin{aligned}
m_{ij}&=2K^{ia}M_{aj}\\
&=2K^{ia}(-K_{aj}+3K_{a}K_{j})\\
&=-2\delta_{ij}-6s_{i}K_{j},
\end{aligned}
\end{equation}
where we use the no scale relationship Eq (\ref{noscale}).

The eigenvalues are:
\begin{equation}
\begin{aligned}
&\lambda_{1}=...\lambda_{N-1}=-2,\\
&\lambda_{N}=-2-6s_{i}K_{i}\\
&=-2-6s_{i}\frac{-3\frac{\partial \mathcal{V}}{\partial s_{i}}}{\mathcal{V}}\\
&=-2+18\times \frac{\frac{7\mathcal{V}}{3}}{\mathcal{V}}\\
&=40,
\end{aligned}
\end{equation}
where we use the property that $\mathcal{V}$ is a homogeneous function of $s_{i}$ of degree $\frac{7}{3}$.\\

Using the standard relationship $\Delta(\Delta-3)=m^{2}R_{AdS}^{2}$, the corresponding conformal dimensions are:
\begin{equation}
\Delta_{1}=...=\Delta_{N-1}=2,
\end{equation}
\begin{equation}
\Delta_{N}=8.
\end{equation}
For the axions, the corresponding Hessian is:
\begin{equation}
M_{t_{i}t_{j}}=2K_{i}K_{j}.
\end{equation}
Similarly, the matrix elements of $m=2K^{-1}M$ are expressed as:
\begin{equation}
\begin{aligned}
m_{ij}&=2K^{ia}M_{aj}\\
&=4K^{ia} K_{a}K_{j}\\
&=-4s_{i}K_{j}.
\end{aligned}
\end{equation}
The corresponding conformal dimensions for the axions are:
\begin{equation}
\Delta_{1}=...=\Delta_{N-1}=3,
\end{equation}
\begin{equation}
\Delta_{N}=7.
\end{equation}

This is rather interesting because all the conformal dimensions are integers. This is reminiscent of DGKT, where similar results were found for general Calabi Yau manifolds \cite{Conlon:2021cjk,Apers:2022tfm}. It was argued in \cite{Kachru:2001je} that for a special class of M-theory compactifications on 7d manifolds X with $G_{2}$ holonomy, special loci in their moduli space are well described by type IIA orientifolds. Therefore, it is natural to compare with DGKT. There may be some connections between the integers of M-Theory and the integers in DGKT, but their values are different. We give a summary of the results for general DGKT.

Following \cite{Apers:2022tfm}, for the  K\" ahler moduli and axion-dilaton sector:
\begin{equation}
\Delta_{1}=10, \quad \Delta_{2...h^{1,1}_{-}+1}=6,
\end{equation}
for the saxions and 
\begin{equation}
\Delta_{1}=11, \quad \Delta_{2...h^{1,1}_{-}+1}=5.
\end{equation}
for the corresponding axions. For the complex structure moduli sector:
\begin{equation}
\Delta_{u_{a}}=2,\quad \Delta_{a_{a}}=3, \quad a=1,...h^{2,1}.
\end{equation}
In DGKT, the K\" ahler potential and superpotential can be expressed as\cite{DeWolfe:2005uu}:
\begin{equation}
K=K^{K}+K^{Q}=-\textrm{Log}(\mathcal{V})+4D,
\end{equation}
\begin{equation}
W=W^{K}+W^{Q},
\end{equation}
where
\begin{equation}
W^{K}=e_{0}+e_{a}z^{a}+\frac{1}{2}k_{abc}m_{a}z_{b}z_{c}-\frac{m_{0}}{6}k_{abc}z_{a}z_{b}z_{c},
\end{equation}
and
\begin{equation}
W^{Q}=-2p_{k}N_{k}-iq_{\lambda}T_{\lambda}.
\end{equation}

Scale separation and geometric limit rely on large $c_{2}\gg 1$. Unlike DGKT, no flux number can be dial up. Note that in DGKT fluxes can be chosen arbitrary large for any Calabi Yau 3-folds. However, in the M-Theory flux-stabilization, the value of $c_{2}$ depends on the topological conditions of the 7d $G_{2}$ manifold. An example of large $c_{2}$ is given in \cite{Acharya:2002kv}, where $Q=H^{3}/\Gamma$.

Note that the complex structure moduli sector in DGKT is similar to M-theory flux vacuum. This is because $W_{Q}$ is also a linear combination of complex structure moduli, which is alone the same line as Eq (\ref{super1}). The resemblance has been predicted in \cite{DeWolfe:2005uu}. However, the volume moduli and axion-dilaton sector seems to be different because there is no quadratic and cubic terms in Eq (\ref{super1}).

\section{Stabilization By Non-Perturbative Effects}
Up to now we have focused on scenarios arising from M-theory flux vacua. In this section, we study a different model based on M-theory. The authors took a further step in \cite{PhysRevLett.97.191601,Acharya:2007rc}: They considered the M-theory vacuum with zero flux background. We will revisit this model and focus on the detailed properties of its holographic dual in this section. 

We give a quick summary to the model. The  K\" ahler potential remains the same in this case:
\begin{equation}
K=-3\textrm{Log}(\mathcal{V}),
\end{equation}
where $\mathcal{V}=\prod_{i=1}^{N}s_{i}^{a_{i}}$ with $\Sigma_{i=1}^{N}a_{i}=\frac{7}{3}.$
The superpotential is generated by the non-perturbative effects\cite{PhysRevLett.97.191601,Acharya:2007rc}:
\begin{equation}
\label{super2}
W=A_{1}e^{i b_{1}\Sigma_{i=1}^{N}N_{i}^{1}z_{i}}+A_{2}e^{i b_{2}\Sigma_{i=1}^{N}N_{i}^{2}z_{i}},
\end{equation}
where $A_{k}$ are numerical constants. $b_{1}=\frac{2\pi}{P}$ and $b_{2}=\frac{2\pi}{Q}$ with $\textit{P}, \textit{Q}$ being the rank of the gauge group for gauge condensation. The sets of $N_{i}^{1}, N_{i}^{2}$ are all integers. Therefore, the M-theory vacuum is fully determined by the constants $(a_{i}, b_{1}, b_{2}, N_{i}^{1}, N_{i}^{2}, A_{1}, A_{2})$. In this paper, without loss of generality, we take positive $A_{1}, A_{2}$.

The SUSY condition is:
\begin{equation}
\label{eq:susy}
D_{i}W=\partial_{i}W+K_{i}W=0.
\end{equation}
The vacuum solution to Eq (\ref{eq:susy}) is:
\begin{equation}
\frac{A_{1}}{A_{2}}=-\textrm{cos}[(b_{1}\Sigma_{i=1}^{N}N^{1}_{i}s_{i}-b_{2}\Sigma_{i=1}^{N}N^{2}_{i}s_{i})]\frac{2b_{2}N^{2}_{i}s_{i}+3a_{i}}{2b_{1}N^{1}_{i}s_{i}+3a_{i}}e^{(b_{1}\Sigma_{i=1}^{N}N^{1}_{i}s_{i}-b_{2}\Sigma_{i=1}^{N}N^{2}_{i}s_{i})},
\end{equation}
\begin{equation}
\textrm{sin}[(b_{1}\Sigma_{i=1}^{N}N^{1}_{i}s_{i}-b_{2}\Sigma_{i=1}^{N}N^{2}_{i}s_{i})]=0.
\end{equation}
To prove them, we can start by taking the expression of $W$ and $K$ into SUSY condition :

\begin{equation}
i(A_{1}b_{1}N^{1}_{i}e^{i b_{1}\Sigma_{i=1}^{N}N^{1}_{i}z_{i}}+A_{2}b_{2}N^{2}_{i}e^{i b_{2}\Sigma_{i=1}^{N}N^{2}_{i}z_{i}})-\frac{3a_{i}}{2is_{i}}(A_{1}e^{i b_{1}\Sigma_{i=1}^{N}N^{1}_{i}z_{i}}+A_{2}e^{i b_{2}\Sigma_{i=1}^{N}N^{2}_{i}z_{i}})=0.
\end{equation}
which implies:
\begin{equation}
\begin{aligned}
\frac{A_{1}}{A_{2}}&=-\frac{2b_{2}N^{2}_{i}s_{i}+3a_{i}}{2b_{1}N^{1}_{i}s_{i}+3a_{i}}e^{ib_{2}\Sigma_{i=1}^{N}N^{2}_{i}z_{i}-i b_{1}\Sigma_{i=1}^{N}N^{1}_{i}z_{i}}\\
&=-\frac{2b_{2}N^{2}_{i}s_{i}+3a_{i}}{2b_{1}N^{1}_{i}s_{i}+3a_{i}}e^{-b_{2}\Sigma_{i=1}^{N}N^{2}_{i}s_{i}+ b_{1}\Sigma_{i=1}^{N}N^{1}_{i}s_{i}}\\
&(\textrm{cos}[-b_{2}\Sigma_{i=1}^{N}N^{2}_{i}t_{i}+ b_{1}\Sigma_{i=1}^{N}N^{1}_{i}t_{i}]-i\textrm{sin}[-b_{2}\Sigma_{i=1}^{N}N^{2}_{i}t_{i}+ b_{1}\Sigma_{i=1}^{N}N^{1}_{i}t_{i}]).\\
\end{aligned}
\end{equation}
So we prove the SUSY condition.

The overall phase of $\textit{W}$,
$e^{ib_{1}\Sigma_{i=1}^{N}N^{1}_{i}t_{i}}$,
does not have physical meaning, since it is only relative phase factors that matter. If we have $A_{1}, A_{2}$ real and positive, the SUSY condition implies:
\begin{equation}
\label{av}
\textrm{cos}[(-b_{2}\Sigma_{i=1}^{N}N^{2}_{i}t_{i}+ b_{1}\Sigma_{i=1}^{N}N^{1}_{i}t_{i})]=-1,
\end{equation}
which also gives:
\begin{equation}
\label{eq:1}
\frac{A_{1}}{A_{2}}=\frac{2b_{2}N^{2}_{i}s_{i}+3a_{i}}{2b_{1}N^{1}_{i}s_{i}+3a_{i}}e^{-b_{2}\Sigma_{i=1}^{N}N^{2}_{i}s_{i}+ b_{1}\Sigma_{i=1}^{N}N^{1}_{i}s_{i}},
\end{equation}
which results in the following equations:
\begin{equation}
\label{16}
\frac{A_{2}}{A_{1}}=\frac{1}{\alpha} e^{-b_{1}\sum_{i=1}^{N}N^{1}_{i}s_{i}+b_{2}\sum_{i=1}^{N}N^{2}_{i}s_{i}},
\end{equation}
\begin{equation}
\label{11}
s_{i}=-\frac{3a_{i}(\alpha-1)}{2(b_{1}N^{1}_{i}\alpha-b_{2}N^{2}_{i})},
\end{equation}
where one can solve $\alpha$ and $s_{i}$ numerically.

The Hessians of the volume moduli and the axions in AdS units read:
\begin{equation}
\begin{aligned}
H^{V}_{ab}=R_{AdS}^{2}\textrm{V}_{ab}=\frac{\textrm{V}_{ab}}{e^{K}|W|^{2}}&=-K_{ab}+3K_{a}K_{b}+2\frac{W_{ab}}{W}+8K^{ij}\frac{W_{ia}W_{jb}}{|W|^{2}}\\
&+2s_{i}K_{b}\frac{W_{ia}}{W}+2s_{j}K_{a}\frac{W_{jb}}{W},\\
\end{aligned}
\end{equation}
and
\begin{equation}
\begin{aligned}
H^{A}_{ab}=R_{AdS}^{2}\textrm{V}_{ab}=\frac{\textrm{V}_{ab}}{e^{K}|W|^{2}}&=2K_{a}K_{b}+6\frac{W_{ab}}{W}+8K^{ij}\frac{W_{ia}W_{jb}}{|W|^{2}}\\
&+2s_{i}K_{b}\frac{W_{ia}}{W}+2s_{j}K_{a}\frac{W_{jb}}{W}.\\
\end{aligned}
\end{equation}
The physical masses can be acquired as the eigenvalues of $M=2K^{-1}H$. Using the explicit expressions for the Hessians in the Appendix, we have:
\begin{equation}
\begin{aligned}
M^{V}_{ab}=2K^{ai}H^{V}_{ib}=&=-2\delta_{ab}+\alpha\frac{s_{a}^{2}}{3a_{a}}\left(b_{1}N^{1}_{a}+\frac{3a_{a}}{2s_{a}}\right)\left(b_{1}N^{1}_{b}+\frac{3a_{b}}{2s_{b}}\right)\left(\Sigma_{i=1}^{N}\frac{8}{3a_{i}}\alpha\left(b_{1}N^{1}_{i}s_{i}+\frac{3a_{i}}{2}\right)^{2}-2\right),\\
\end{aligned}
\end{equation}
\begin{equation}
\begin{aligned}
M^{A}_{ab}=2K^{ai}H^{A}_{ib}=&=\alpha\frac{s_{a}^{2}}{3a_{a}}\left(b_{1}N^{1}_{a}+\frac{3a_{a}}{2s_{a}}\right)\left(b_{1}N^{1}_{b}+\frac{3a_{b}}{2s_{b}}\right)\left(\Sigma_{i=1}^{N}\frac{8}{3a_{i}}\alpha\left(b_{1}N^{1}_{i}s_{i}+\frac{3a_{i}}{2}\right)^{2}-6\right).\\
\end{aligned}
\end{equation}
The eigenvalues are:
\begin{equation}
\lambda^{V}_{i}=-2, \quad i=1,...N-1,
\end{equation}
\begin{equation}
\begin{aligned}
\lambda^{V}_{N}&=-2+\alpha\Sigma_{a=1}^{N}\frac{s_{a}^{2}}{3a_{a}}\left(b_{1}N^{1}_{a}+\frac{3a_{a}}{2s_{a}}\right)^{2}\left(\Sigma_{i=1}^{N}\frac{8}{3a_{i}}\alpha\left(b_{1}N^{1}_{i}s_{i}+\frac{3a_{i}}{2}\right)^{2}-2\right)\\
&=-2+\alpha\Sigma_{a=1}^{N}\frac{1}{3a_{a}}\left(b_{1}N^{1}_{a}s_{a}+\frac{3a_{a}}{2}\right)^{2}\left(\Sigma_{i=1}^{N}\frac{8}{3a_{i}}\alpha\left(b_{1}N^{1}_{i}s_{i}+\frac{3a_{i}}{2}\right)^{2}-2\right),\\
\end{aligned}
\end{equation}
and
\begin{equation}
\lambda^{A}_{i}=0,\quad i=1,...N-1,
\end{equation}
\begin{equation}
\begin{aligned}
\lambda^{A}_{N}&=\alpha\Sigma_{a=1}^{N}\frac{s_{a}^{2}}{3a_{a}}\left(b_{1}N^{1}_{a}+\frac{3a_{a}}{2s_{a}}\right)^{2}\left(\Sigma_{i=1}^{N}\frac{8}{3a_{i}}\alpha\left(b_{1}N^{1}_{i}s_{i}+\frac{3a_{i}}{2}\right)^{2}-6\right)\\
&=\alpha\Sigma_{a=1}^{N}\frac{1}{3a_{a}}\left(b_{1}N^{1}_{a}s_{a}+\frac{3a_{a}}{2}\right)^{2}\left(\Sigma_{i=1}^{N}\frac{8}{3a_{i}}\alpha\left(b_{1}N^{1}_{i}s_{i}+\frac{3a_{i}}{2}\right)^{2}-6\right).\\
\end{aligned}
\end{equation}
respectively.\\
Using the standard relationship $\Delta(\Delta-3)=m^{2}R^{2}_{AdS}$,
the corresponding conformal dimensions are:
\begin{equation}
\Delta^{V}_{i}=2, \quad i=1,...N-1,
\end{equation}
\begin{equation}
\label{cd}
\Delta^{V}_{N}=\frac{3+\sqrt{1+4\alpha\Sigma_{a=1}^{N}\frac{1}{3a_{a}}\left(b_{1}N^{1}_{a}s_{a}+\frac{3a_{a}}{2}\right)^{2}\left(\Sigma_{i=1}^{N}\frac{8}{3a_{i}}\alpha\left(b_{1}N^{1}_{i}s_{i}+\frac{3a_{i}}{2}\right)^{2}-2\right)}}{2},
\end{equation}
and
\begin{equation}
\Delta^{A}_{i}=3, \quad i=1,...N-1,
\end{equation}
\begin{equation}
\Delta^{A}_{N}=\frac{3+\sqrt{9+4\alpha\Sigma_{a=1}^{N}\frac{1}{3a_{a}}\left(b_{1}N^{1}_{a}s_{a}+\frac{3a_{a}}{2}\right)^{2}\left(\Sigma_{i=1}^{N}\frac{8}{3a_{i}}\alpha\left(b_{1}N^{1}_{i}s_{i}+\frac{3a_{i}}{2}\right)^{2}-6\right)}}{2}.
\end{equation}
Note that the option $\Delta^{V}_{i}=1$ is excluded by $N=1$ supersymmetry, as the volume moduli and the axions are in the same 3d $N=1$ supermultiplet \cite{Cordova:2016emh}.

Following \cite{Acharya:2007rc}, we work in the following two branches in which the supergravity description is meaningful:
\begin{equation}
\begin{aligned}
&a)\frac{A_{2}}{A_{1}}>1, \textrm{min}{\frac{b_{2}N_{i}^{2}}{b_{1}N_{i}^{1}};i=1...N}>\alpha>\textrm{max}{\frac{b_{2}N_{i}^{2}+\frac{3a_{i}}{2}}{b_{1}N_{i}^{1}+\frac{3a_{i}}{2}};i=1...N}\\
&b)\frac{A_{2}}{A_{1}}<1, \textrm{max}{\frac{b_{2}N_{i}^{2}}{b_{1}N_{i}^{1}};i=1...N}<\alpha<\textrm{min}{\frac{b_{2}N_{i}^{2}+\frac{3a_{i}}{2}}{b_{1}N_{i}^{1}+\frac{3a_{i}}{2}};i=1...N}\\
\end{aligned}
\end{equation}

It remains to show how the conformal dimension (\ref{cd}) grows with the $R_{AdS}$ in the scale seperation limit where $R_{AdS}$ goes to infinity. Since we have a supersymmetric vacuum, it follows that $V=-3e^{K}|W|^{2}$, therefore
$\frac{1}{R_{AdS}^{2}}=e^{K}|W|^{2}$,
which implies:
\begin{equation}
\textrm{Log}  R_{AdS}=-\frac{1}{2}K- \textrm{Log} W.
\end{equation}
Following the definition of  K\" ahler potential, superpotential, Eq (\ref{16}), Eq (\ref{18}) and combining them, we have:
\begin{equation}
\textrm{Log}   R_{AdS}=\Sigma_{i=1}^{N}\frac{3}{2}a_{i}\textrm{Log} s_{i}+b_{1}\Sigma N_{i}^{1}s_{i}+\textrm{Log}\left(A_{1}\left(1-\frac{1}{\alpha}\right)\right).
\end{equation}
In the scale separation limit, the leading order contribution of the conformal dimension and the $\textrm{Log}R_{AdS}$ is:
\begin{equation}
\Delta_{N}\propto \alpha\Sigma_{a=1}^{N}\frac{1}{3a_{a}}\left(b_{1}N^{1}_{a}s_{a}+\frac{3a_{a}}{2}\right)^{2}\propto b_{1}(\Sigma_{i=1}^{N} N_{i}^{1}s_{i})^{2},
\end{equation}
\begin{equation}
\textrm{Log}^{2}R_{AdS}\propto b_{1}(\Sigma_{i=1}^{N} N_{i}^{1}s_{i})^{2},
\end{equation}
Therefore, in the scale separation limit:
\begin{equation}
\Delta_{Heavy}\propto \textrm{Log}^{2}R_{AdS}.
\end{equation}

We end this section by comparing the results with other scenarios like KKLT \cite{Kachru:2003aw} and Racetrack \cite{Escoda:2003fa}, as they are both non-perturbative stabilized. KKLT is a type IIB flux compactification model in which both fluxes and non-perturbative effects are used to stabilize the moduli supersymmetrically. The AdS minimum is found after moduli stabilization then lifted to dS. The 4d effective field theory is described by:
\begin{equation}
\begin{aligned}
&K=-3\textrm{Log}(-i(\rho-\overline{\rho})),\\
&W=W_{0}+Ae^{i\alpha \rho}.\\
\end{aligned}
\end{equation}
In the limit when $|W_{0}|\ll 1$,  one obtains $\Delta\propto \textrm{Log} R_{AdS}$ \cite{Conlon:2018vov,Choi:2005ge}.

Racetrack still occurs in the context of both type IIB and heterotic string compactification but with only non-perturbative effects. The K\" ahler potential is the same as KKLT while the superpotential are dominated by two different non-perturbative effects:
\begin{equation}
W=Ae^{i\alpha \rho}-Be^{i\beta \rho}.
\end{equation}
Following \cite{Conlon:2020wmc}, one can show  $\Delta\propto \textrm{Log}^{2} R_{AdS}$ in the scale separation limit. 

It is natural that M-theory gives similar results to IIB racetrack because racetrack model corresponds to the one modulus case of M-theory with zero flux background. 
\section{Conclusion}

We finish this paper by summarizing the results and proposing open questions. In this paper, we study two different models of M-theory moduli stabilization.

First, for the flux-stabilized M-Theory vacuum, we have shown that the spectrum of the dual $CFT_{3}$ is chacterized by a set of integer conformal dimensions. The presence of integer conformal dimensions dual to the moduli and axions in M-Theory flux vacuum is quite intriguing. Similar results also appear in DGKT type IIA string compactification scenarios\cite{Conlon:2021cjk, Apers:2022tfm}. Therefore we compared them and found the flux-stabilized M-Theory vacuum resembles the complex structure sector of DGKT, which is not surprising because their superpotential has similar forms, as predicted in \cite{DeWolfe:2005uu}. The validity of the results heavily rely on the existence of large $c_{2}$.

Second, for non-perturbatively effects stabilized M-Theory vacuum, we prove that there is one heavy operator with $\Delta\propto\textrm{Log} R^{2}_{AdS}$, the rest are the same as flux-stabilized M-Theory vacuum. Then we compare the results with other scenarios like KKLT \cite{Kachru:2003aw} and racetrack \cite{Escoda:2003fa}. It is different with KKLT because there are two different non-perturbative effects in the superpotential, while KKLT only has one non-perturbative term. Racetrack is quite similar because it is the one modulus version of the non-perturbative effects stabilized M-Theory vacuum.

An extremely interesting question is: What is the origin of these integer conformal dimensions? Currently we have no explanation for this, but we believe there must be some deeper structures behind this. Another interesting open question is: Can we prove the existence of $G_{2}$ manifold with large $c_{2}$? It would certainly be of interest to explore these questions in the future.

\section*{Acknowledgements}

I want to thank Joseph Conlon for suggesting this project and his continuous guidance during the project. I am also very grateful to Joseph Conlon for many fruitful discussions and
helpful comments on the manuscript. I want to thank Fien Apers and Filippo Revello for comments on the manuscript. I acknowledges funding support from the China Scholarship Council-FaZheng Group- University of Oxford. 

\appendix
\section{General Expressions For Mass Matrices}\label{a1}
In this section we derive Eq (\ref{eq:3}) and Eq (\ref{eq:4}).
The general expressions of moduli and axions Hessians are:
\begin{equation}
\begin{aligned}
\label{eq:mass matrix1}
H_{ab}^{V}=\partial_{b}\partial_{a}\textrm{V}&=K_{ab}\textrm{V}-3e^{K}K_{b}\partial_{a}|W|^{2}-3e^{K}\partial_{b}\partial_{a}|W|^{2}\\
&+e^{K}K^{i\overline{j}}(\partial_{a}D_{i}W)(\partial_{b}D_{\overline{j}}\overline{W})+e^{K}K^{i\overline{j}}(\partial_{b}D_{i}W)(\partial_{a}D_{\overline{j}}\overline{W}),\\
\end{aligned}
\end{equation}
where $\partial_{a}$,$\partial_{b}$ means $\partial_{s_{a}}$, $\partial_{s_{b}}$ and $\partial_{i}$,$\partial_{j}$ means $\partial_{z_{i}}$, $\partial_{z_{j}}$.
\begin{equation}
\begin{aligned}
\label{eq:mass matrix2}
H_{ab}^{A}=\partial_{b}\partial_{a}V&=-3e^{K}\partial_{b}\partial_{a}|W|^{2}+e^{K}K^{i\overline{j}}(\partial_{a}D_{i}W)(\partial_{b}D_{\overline{j}}\overline{W})\\
&+e^{K}K^{i\overline{j}}(\partial_{b}D_{i}W)(\partial_{a}D_{\overline{j}}\overline{W}),\\
\end{aligned}
\end{equation}
where $\partial_{a}$,$\partial_{b}$ means $\partial_{t_{a}}$, $\partial_{t_{b}}$ and $\partial_{i}$,$\partial_{j}$ means $\partial_{z_{i}}$, $\partial_{z_{j}}$.

For the first line of (\ref{eq:mass matrix1}), the SUSY condition implies that:
\begin{equation}
\label{app1}
\partial_{a}|W|^{2}=2\partial_{a}W\overline{W}=-K_{a}|W|^{2},
\end{equation}
and
\begin{equation}
\begin{aligned}
\label{app2}
&\partial_{b}\partial_{a}|W|^{2}=2\partial_{b}\partial_{a}W\overline{W}+2\partial_{a}W\partial_{b}\overline{W}\\
&=(2\frac{W_{ab}}{W}+\frac{1}{2}K_{a}K_{b})|W|^{2}.\\
\end{aligned}
\end{equation}
Following \cite{Apers:2022tfm}, for the second line of (\ref{eq:mass matrix1}), we have:
\begin{equation}
\begin{aligned}
&e^{K}K^{i\overline{j}}(\partial_{a}D_{i}W)(\partial_{b}D_{\overline{j}}\overline{W})\\
&=e^{K}(K^{i\overline{j}}W_{ia}\overline{W}_{\overline{j}b}+4W_{ab}\overline{W}-\frac{1}{2}K^{i\overline{j}}W_{ia}K_{\overline{j}}K_{b}\overline{W}\\
&-\frac{1}{2}K^{i\overline{j}}W_{ib}K_{\overline{j}}K_{a}\overline{W}+(K_{ab}+\frac{3}{4}K_{a}K_{b})|W|^{2}).\\
\end{aligned}
\end{equation}
The fact that $K=K(z_{i}+\overline{z}_{i})$ and the superpotential is holomorphic implies:
\begin{equation}
\partial_{z_{i}}K=\frac{\partial_{s_{i}}K}{2i}, \partial_{\overline{z}_{i}}K=-\frac{\partial_{s_{i}}K}{2i},
\end{equation}
and
\begin{equation}
\partial_{z_{i}}W=\frac{1}{i}\partial_{s_{i}}W, \partial_{\overline{z}_{i}}\overline{W}=-\frac{1}{i}\partial_{s_{i}}\overline{W},
\end{equation}
which results in:
\begin{equation}
\begin{aligned}
\label{app3}
&e^{K}K^{i\overline{j}}(\partial_{a}D_{i}W)(\partial_{b}D_{\overline{j}}\overline{W})\\
&=e^{K}(4K^{ij}W_{ia}\overline{W}_{jb}+4W_{ab}\overline{W}+s_{i}W_{ia}K_{b}\overline{W}\\
&+s_{j}\overline{W}_{jb}K_{a}W+(K_{ab}+\frac{3}{4}K_{a}K_{b})|W|^{2}).\\
\end{aligned}
\end{equation}
Combining (\ref{app1}),(\ref{app2}) and (\ref{app3}) leads to:
\begin{equation}
\begin{aligned}
M^{V}_{ab}=\frac{\textrm{V}_{ab}}{e^{K}|W|^{2}}&=-K_{ab}+3K_{a}K_{b}+2\frac{W_{ab}}{W}+8K^{ij}\frac{W_{ia}W_{jb}}{|W|^{2}}\\
&+2s_{i}K_{b}\frac{W_{ia}}{W}+2s_{j}K_{a}\frac{W_{jb}}{W},\\
\end{aligned}
\end{equation}
The axion Hessians (\ref{eq:mass matrix2}) can be obtained similarly, note that:
\begin{equation}
\label{m2}
\frac{\partial W}{\partial t_{a}}=\frac{\partial W}{\partial z_{a}}=\frac{1}{i}\frac{\partial W}{\partial s_{a}},
\end{equation}
and
\begin{equation}
\begin{aligned}
\label{m23}
\partial_{a}D_{i}W&=\partial_{a}(\partial_{z_{i}}W+K_{i}W)\\
&=\partial_{z_{i}}\partial_{z_{a}}W+K_{z_{i}}\partial_{z_{a}}W\\
&=W_{z_{i}z_{a}}-K_{z_{i}}K_{z_{a}}W\\
&=-W_{ia}+\frac{1}{4}K_{i}K_{a}W
\end{aligned}
\end{equation}
which results in:
\begin{equation}
\label{ax1}
\begin{aligned}
\partial_{b}\partial_{a}|W|^{2}&=2\partial_{t_{b}}\partial_{t_{a}}W\overline{W}+2\partial_{t_{a}}W\partial_{t_{b}}\overline{W}\\
&=-2W_{ab}\overline{W}+\frac{1}{2}K_{a}K_{b}|W|^{2}.\\
\end{aligned}
\end{equation}
For the second term of (\ref{eq:mass matrix2}), using (\ref{m2}), (\ref{m23}) we have:
\begin{equation}
\label{ax2}
\begin{aligned}
&e^{K}K^{i\overline{j}}(\partial_{a}D_{i}W)(\partial_{b}D_{\overline{j}}\overline{W})\\
&=e^{K}4K^{ij}(-W_{ia}+\frac{1}{4}K_{i}K_{a}W)(-\overline{W}_{jb}+\frac{1}{4}K_{j}K_{b}\overline{W})\\
&=e^{K}(4K^{ij}W_{ia}W_{jb}+s_{i}W_{ia}K_{b}W+s_{j}W_{jb}K_{a}W+\frac{7}{4}K_{a}K_{b}|W|^{2}).\\
\end{aligned}
\end{equation}
Putting (\ref{ax1}), (\ref{ax2}) together into (\ref{eq:mass matrix2}):
\begin{equation}
\begin{aligned}
M^{A}_{ab}=\frac{\textrm{V}_{ab}}{e^{K}|W|^{2}}&=2K_{a}K_{b}+6\frac{W_{ab}}{W}+8K^{ij}\frac{W_{ia}W_{jb}}{|W|^{2}}\\
&+2s_{i}K_{b}\frac{W_{ia}}{W}+2s_{j}K_{a}\frac{W_{jb}}{W}.\\
\end{aligned}
\end{equation}
\section{Hessians Of Zero Fluxes Background} 
In this appendix we give the explicit expressions for the Hessians of the moduli and axions.\\
We can subsititute the vaccum expectation value of axions Eqs(\ref{av}) into the superpotential because the Hessians for moduli and axions factorise:
\begin{equation}
\label{18}
W=A_{1}e^{- b_{1}\Sigma_{i=1}^{N}N_{i}^{1}s_{i}}-A_{2}e^{-b_{2}\Sigma_{i=1}^{N}N_{i}^{2}s_{i}}.
\end{equation}
The first and second derivatives for $\textit{W}$ are:
\begin{equation}
W_{a}=-b_{1}N^{1}_{a}A_{1}e^{-b_{1}\sum_{i=1}^{N}N^{1}_{i}s_{i}}+b_{2}N^{2}_{a}A_{2}e^{-b_{2}\sum_{i=1}^{N}N^{2}_{i}s_{i}},
\end{equation}
\begin{equation}
\label{81}
W_{ab}=b_{1}^{2}N^{1}_{a}N^{1}_{b}A_{1}e^{-b_{1}\sum_{i=1}^{N}N^{1}_{i}s_{i}}-b_{2}^{2}N^{2}_{a}N^{2}_{b}A_{2}e^{-b_{2}\sum_{i=1}^{N}N^{2}_{i}s_{i}}.
\end{equation}
We substitute the SUSY condition (\ref{eq:1}) into (\ref{81}):
\begin{equation}
\frac{W_{ab}}{W}=\frac{\alpha b_{1}^{2}N^{1}_{a}N^{1}_{b}-b_{2}^{2}N^{2}_{a}N^{2}_{b}}{\alpha-1}.
\end{equation}
Eqs(\ref{11}) implies:
\begin{equation}
N_{a}^{2}=\frac{b_{1}N_{a}^{1}}{b_{2}}\alpha+\frac{3a_{a}(\alpha-1)}{2s_{a}b_{2}}.
\end{equation}
Putting them together, the second order derivatives of the superpotential are expressed as:
\begin{equation}
\begin{aligned}
\frac{W_{ab}}{W}&=-\alpha b_{1}^{2}N_{a}^{1}N_{b}^{1}-\frac{3}{2}b_{1}\alpha\left(\frac{a_{b}}{s_{b}}N_{a}^{1}+\frac{a_{a}}{s_{a}}N_{b}^{1}\right)+\frac{9a_{a}a_{b}}{4s_{a}s_{b}}\left(1-\alpha\right)\\
&=-\alpha\left(b_{1}N^{1}_{a}+\frac{3a_{a}}{2s_{a}}\right)\left(b_{1}N^{1}_{b}+\frac{3a_{b}}{2s_{b}}\right)+\frac{9a_{a}a_{b}}{4s_{a}s_{b}},\\
\end{aligned}
\end{equation}

The Hessians of the moduli and axions can be expressed in AdS units:
\begin{equation}
\begin{aligned}
\label{15}
H^{V}_{ab}=R_{AdS}^{2}\textrm{V}_{ab}=\frac{\textrm{V}_{ab}}{e^{K}|W|^{2}}&=-K_{ab}+3K_{a}K_{b}+2\frac{W_{ab}}{W}+8K^{ij}\frac{W_{ia}W_{jb}}{|W|^{2}}\\
&+2s_{i}K_{b}\frac{W_{ia}}{W}+2s_{j}K_{a}\frac{W_{jb}}{W}.\\
\end{aligned}
\end{equation}
\begin{equation}
\begin{aligned}
H^{A}_{ab}=R_{AdS}^{2}\textrm{V}_{ab}=\frac{\textrm{V}_{ab}}{e^{K}|W|^{2}}&=2K_{a}K_{b}+6\frac{W_{ab}}{W}+8K^{ij}\frac{W_{ia}W_{jb}}{|W|^{2}}\\
&+2s_{i}K_{b}\frac{W_{ia}}{W}+2s_{j}K_{a}\frac{W_{jb}}{W}.\\
\end{aligned}
\end{equation}
The derivatives of K\" ahler potential and superpotential are given by:
\begin{equation}
-K_{ab}=-\frac{s_{a}^{2}}{3a_{a}}\delta_{ab},
\end{equation}
\begin{equation}
K_{a}K_{b}=\frac{9a_{a}a_{b}}{s_{a}s_{b}},
\end{equation}
\begin{equation}
2\frac{W_{ab}}{W}=-2\alpha\left(b_{1}N^{1}_{a}+\frac{3a_{a}}{2s_{a}}\right)\left(b_{1}N^{1}_{b}+\frac{3a_{b}}{2s_{b}}\right)+\frac{9a_{a}a_{b}}{2s_{a}s_{b}}.
\end{equation}
These are used to evaluate the last term in the first line in (\ref{15}):
\begin{equation}
\begin{aligned}
8K^{ij}\frac{W_{ia}W_{jb}}{|W|^{2}}&=8\Sigma_{i=1}^{N}\frac{s_{i}^{2}}{3a_{i}}\left(-\alpha\left(b_{1}N^{1}_{a}+\frac{3a_{a}}{2s_{a}}\right)+\frac{9a_{i}a_{a}}{4s_{i}s_{a}}\right)\left(-\alpha\left(b_{1}N^{1}_{i}+\frac{3a_{i}}{2s_{i}}\right)\left(b_{1}N^{1}_{b}+\frac{3a_{b}}{2s_{b}}\right)+\frac{9a_{i}a_{b}}{4s_{i}s_{b}}\right)\\
&=\frac{8}{3a_{i}}\alpha^{2}\left(b_{1}N^{1}_{i}s_{i}+\frac{3a_{i}}{2}\right)^{2}\left(b_{1}N^{1}_{a}+\frac{3a_{a}}{2s_{a}}\right)\left(b_{1}N^{1}_{b}+\frac{3a_{b}}{2s_{b}}\right)\\
&-\frac{8s_{i}^{2}}{3a_{i}}\alpha \left(b_{1}N_{i}^{1}+\frac{3a_{i}}{2s_{i}}\right)\frac{9a_{i}}{4s_{i}}\left(\left(b_{1}N^{1}_{a}
+\frac{3a_{a}}{2s_{a}}\right)\frac{a_{b}}{s_{b}}+\left(b_{1}N^{1}_{b}
+\frac{3a_{b}}{2s_{b}}\right)\frac{a_{a}}{s_{a}}\right)+\frac{8s_{i}^{2}}{3a_{i}}\frac{81a_{i}^{2}a_{a}a_{b}}{16s_{i}^{2}s_{a}s_{b}}\\
=&\frac{8}{3a_{i}}\alpha^{2}\left(b_{1}N^{1}_{i}s_{i}+\frac{3a_{i}}{2}\right)^{2}\left(b_{1}N^{1}_{a}+\frac{3a_{a}}{2s_{a}}\right)\left(b_{1}N^{1}_{b}+\frac{3a_{b}}{2s_{b}}\right)\\
&-6s_{i}\alpha \left(b_{1}N_{i}^{1}+\frac{3a_{i}}{2s_{i}}\right)\left(b_{1}N^{1}_{a}\frac{a_{b}}{s_{b}}+b_{1}N^{1}_{b}\frac{a_{a}}{s_{a}}
+\frac{3a_{b}a_{a}}{s_{b}s_{a}}\right)+\frac{63}{2}\frac{a_{a}a_{b}}{s_{a}s_{b}}\\
\end{aligned}
\end{equation}
The second line of (\ref{15}) can be obtained similarly:
\begin{equation}
\begin{aligned}
2s_{i}K_{b}\frac{W_{ia}}{W}&=2s_{i}\frac{-3a_{b}}{s_{b}}\frac{W_{ia}}{W}\\
&=-\frac{6a_{b}}{s_{b}}\Sigma_{i=1}^{N}\left(-\alpha\left(b_{1}N^{1}_{i}s_{i}+\frac{3}{2}a_{i}\right)\left(b_{1}N^{1}_{a}+\frac{3a_{a}}{2s_{a}}\right)+\frac{9a_{a}a_{i}}{4s_{a}}\right)\\
&=\frac{6a_{b}}{s_{b}}\Sigma_{i=1}^{N}\alpha\left(b_{1}N^{1}_{i}s_{i}+\frac{3}{2}a_{i}\right)\left(b_{1}N^{1}_{a}+\frac{3a_{a}}{2s_{a}}\right)-\frac{63a_{a}a_{b}}{2s_{a}s_{b}},
\end{aligned}
\end{equation}
\begin{equation}
\begin{aligned}
2s_{i}K_{a}\frac{W_{ib}}{W}&=2s_{i}\frac{-3a_{a}}{s_{a}}\frac{W_{ib}}{W}\\
&=\frac{6a_{a}}{s_{a}}\Sigma_{i=1}^{N}\alpha\left(b_{1}N^{1}_{i}s_{i}+\frac{3}{2}a_{i}\right)\left(b_{1}N^{1}_{b}+\frac{3a_{b}}{2s_{b}}\right)-\frac{63a_{a}a_{b}}{2s_{a}s_{b}},
\end{aligned}
\end{equation}
Combining these results in:

\begin{equation}
\begin{aligned}
H^{V}_{ab}&=-K_{ab}+3K_{a}K_{b}+2\frac{W_{ab}}{W}+8K^{ij}\frac{W_{ia}W_{jb}}{|W|^{2}}\\
&+2s_{i}K_{b}\frac{W_{ia}}{W}+2s_{j}K_{a}\frac{W_{jb}}{W}.\\
&=-\frac{s_{a}^{2}}{3a_{a}}\delta_{ab}+\frac{8}{3a_{i}}\alpha^{2}\left(b_{1}N^{1}_{i}s_{i}+\frac{3a_{i}}{2}\right)^{2}\left(b_{1}N^{1}_{a}+\frac{3a_{a}}{2s_{a}}\right)\left(b_{1}N^{1}_{b}+\frac{3a_{b}}{2s_{b}}\right)\\
&-2\alpha\left(b_{1}N^{1}_{a}+\frac{3a_{a}}{2s_{a}}\right)\left(b_{1}N^{1}_{b}+\frac{3a_{b}}{2s_{b}}\right)\\
&=-\frac{s_{a}^{2}}{3a_{a}}\delta_{ab}+\alpha\left(b_{1}N^{1}_{a}+\frac{3a_{a}}{2s_{a}}\right)\left(b_{1}N^{1}_{b}+\frac{3a_{b}}{2s_{b}}\right)\left(\frac{8}{3a_{i}}\alpha\left(b_{1}N^{1}_{i}s_{i}+\frac{3a_{i}}{2}\right)^{2}-2\right),\\
\end{aligned}
\end{equation}
and
\begin{equation}
\begin{aligned}
H^{A}_{ab}&=2K_{a}K_{b}+6\frac{W_{ab}}{W}+8K^{ij}\frac{W_{ia}W_{jb}}{|W|^{2}}\\
&+2s_{i}K_{b}\frac{W_{ia}}{W}+2s_{j}K_{a}\frac{W_{jb}}{W}.\\
&=\frac{8}{3a_{i}}\alpha^{2}\left(b_{1}N^{1}_{i}s_{i}+\frac{3a_{i}}{2}\right)^{2}\left(b_{1}N^{1}_{a}+\frac{3a_{a}}{2s_{a}}\right)\left(b_{1}N^{1}_{b}+\frac{3a_{b}}{2s_{b}}\right)\\
&-6\alpha\left(b_{1}N^{1}_{a}+\frac{3a_{a}}{2s_{a}}\right)\left(b_{1}N^{1}_{b}+\frac{3a_{b}}{2s_{b}}\right)\\
&=\alpha\left(b_{1}N^{1}_{a}+\frac{3a_{a}}{2s_{a}}\right)\left(b_{1}N^{1}_{b}+\frac{3a_{b}}{2s_{b}}\right)\left(\frac{8}{3a_{i}}\alpha\left(b_{1}N^{1}_{i}s_{i}+\frac{3a_{i}}{2}\right)^{2}-6\right)\\
\end{aligned}
\end{equation}
\bibliography{bibabstract1}
\bibliographystyle{JHEP}
\end{document}